\documentclass[aps,prb,showpacs,raggedbottom,nobalancelastpage,amssymb,twocolumn,groupedaddress]{revtex4}
\usepackage{graphicx}
\usepackage{amsmath}
\usepackage{amsfonts}
\usepackage{amssymb}
\usepackage{epsfig}
\usepackage{color}
\usepackage{dsfont}

\newcommand{\abs}[1]{\left\vert#1\right\vert}

\newcommand{\bra}[1]{\left\langle#1\right\vert}
\newcommand{\ket}[1]{\left\vert#1\right\rangle}

\begin{document}

\title{A Spin $1/2$ Fermions Chain with $XY$ Interaction}

\author{Gianluca~Francica}
\email{gianluca.francica@gmail.com}

\date{\today}

\begin{abstract}
We consider a chain of spinful fermions with nearest neighbor hopping in the presence of a $XY$ antiferromagnetic interaction. The $XY$ term is mapped onto a Kitaev chain at half-filling such that displays a bosonic zero mode topologically protected and long-range order. As the strength of the hopping amplitude is changed, the system undergoes a quantum phase transition from the topological non-trivial to the trivial phase. We apply the finite-size scaling method to determinate the phase diagram of the model.
\end{abstract}

\maketitle

\section{\label{sec.intro}Introduction}
In the last decades topological materials have received a great attention. Although a systematic study can be performed for non-interacting fermions~\cite{schnyder08,kitaev08,chiu16}, the role of interactions remains particularly attractive because it leads to a breakdown of this classification~\cite{Fidowski10}. In one dimensional interacting systems there is a general framework for classifying
gapped topologically symmetry protected phases, that of extensions of the symmetry group of the model with $U(1)$~\cite{Fidowski11}, or by examining the entanglement spectrum~\cite{Pollmann10,Turner11}.

Among the topological models, the Kitaev chain is a paradigmatic model which displays two Majorana zero modes at the ends in its non-trivial topological phase~\cite{kitaev01}.   Majorana  modes are topologically protected, and are particularly attractive  for  realizing  quantum  registers  which  are  immune to decoherence effects, with promising applications in fault-tolerant quantum computation~\cite{nayak08}.
There have been several proposals and realizations of this model, for example, by using a semiconducting nanowire with strong spin-orbit coupling, proximity-coupled to standard s-wave superconductors and in the presence of a magnetic field~\cite{sau10,alicea10,Lutchyn10,Oreg10}, or alternatively by using ferromagnetic metallic chains~\cite{Nadj-Perge14}.

In this paper, we consider a spin $1/2$ fermions chain with nearest neighbor hopping in the presence of a $XY$ antiferromagnetic interaction favoring a spontaneous magnetization and long-range order for $\gamma\neq 0$, where $\gamma$ is the anisotropy parameter. The $XY$ interaction is mapped onto a Kitaev chain at the half-filling by performing two Jordan-Wigner~\cite{lieb61} and a Mattis-Nam trasformation~\cite{Mattis72}. We show that the $XY$ term displays a bosonic zero mode topologically protected by the symmetry $P_\uparrow$, which is a parity transformation of only the spin up species of the fermions.
In the presence of nearest neighbor hopping the ground state is calculated by a matrix product state variational method~\cite{schollwock11}, and a phase diagram is obtained through the finite size scaling method~\cite{um07}.

The paper is structured in the following way. In the Sect.~\ref{sec.model} we introduce the model. The Sect.~\ref{sec.mapping} is devoted to the mapping onto the Kitaev chain and the discussion of the topological feautures. In Sec.~\ref{sec.qpt} we characterize the quantum phases giving a phase diagram. In Sec.~\ref{sec.conclusion} we summarize the results achieved.

\section{\label{sec.model} Model}
We consider  a one-dimensional chain of  spin $1/2$ fermions described by the Hamiltonian $H=H_0+H_1$.

For a chain  of length $L$  with open boundary conditions, the Hamiltonian $H_0$  reads
\begin{equation}
H_0 = J \sum_{j=1}^{L-1} (1+\gamma) S^x_j S^x_{j+1}+(1-\gamma) S^y_j S^y_{j+1}
\end{equation}
\noindent describing an electronic system with an anti-ferromagnetic $XY$ interaction, where $\gamma$ characterizes the degree of anisotropy in the $xy$-plane. We have introduced the spin operators $S^\alpha_j$ with $\alpha=x,y,z$, which read $S^\alpha_j=\frac{1}{2} c^\dagger_{j \sigma} \sigma^\alpha_{\sigma \sigma'} c_{j \sigma'} $ where $\sigma^\alpha$ are the Pauli matrices and the operators $c_{j\sigma}$ ($c^\dagger_{j\sigma}$) annihilate (create) a fermion on site $j$ with spin $\sigma=\uparrow,\downarrow$ and satisfy the anticommutation relations $\{c_{i\sigma},c_{j\sigma'}\}=0$ and $\{c^\dagger_{i\sigma},c_{j\sigma'}\}=\delta_{ij}\delta_{\sigma \sigma'}$.

The term $H_1$ reads
\begin{equation}
H_1 = -t \sum_{j=1,\sigma}^{L-1} (c^\dagger_{j \sigma} c_{j+1 \sigma} + h.c. )
\end{equation}
\noindent and gives the nearest neighbor hopping.

We note that the model is symmetric with respect to the number parity transformation represented by the unitary operator $P=P_\uparrow P_\downarrow$ where $P_\sigma=e^{i \pi \sum_{j=1}^L n_{j\sigma}}$ and $n_{j \sigma}=c^\dagger_{j\sigma}c_{j \sigma}$ is the occupation number operator. In particular, since the interactions involve products of an even number of fermions of the same spin species, the model is also invariant under the two transformations $P_\sigma$. Other symmetries, like the time-reversal symmetry, are not important for our discussion, i.e. can be broken. Specifically, the symmetry $P_\sigma$ protects a topological phase, appearing for $J$ strong enough and $\gamma\neq 0$ and showing a bosonic zero mode in the thermodynamic limit $L\to \infty$.

Because of the presence of the quartic terms, the Hamiltonian $H$ cannot be diagonalized in straightforward way, also if both the terms $H_0$ and $H_1$ can be individually diagonalized.
Lets show how the Hamiltonian $H_0$ can be mapped in a Kitaev chain at half filling.

\section{\label{sec.mapping} Mapping }
We introduce the Majorana fermion representation by defining the real Majorana operators $a_{ j\sigma} = c_{ j\sigma} + c^\dagger_{ j\sigma} $ and $b_{ j\sigma} = -i c_{ j\sigma} + i c^\dagger_{ j\sigma}$ which satisfy relations $\{a_{i \sigma},a_{j \sigma'}\}= \{b_{i \sigma},b_{j \sigma'}\}=2 \delta_{ij}\delta_{\sigma \sigma'}$ and $\{a_{i \sigma},b_{j \sigma'}\}= 0$.

In this representation the spin operators $S^\alpha_j$ read
\begin{eqnarray}
S^x_j &=& \frac{i}{4}(a_{j\uparrow}b_{j\downarrow}+a_{j\downarrow}b_{j\uparrow})\\
S^y_j &=& -\frac{i}{4}(a_{j\uparrow}a_{j\downarrow}+b_{j\downarrow}b_{j\uparrow})\\
S^z_j &=& \frac{i}{4}(a_{j\uparrow}b_{j\uparrow}+a_{j\downarrow}b_{j\downarrow})
\end{eqnarray}
\noindent By performing the Jordan-Wigner transformation
\begin{eqnarray}
a_{ j \uparrow} &=& \left(\prod_{i=1}^{j-1} \sigma^x_{2 i-1}\right)\sigma^z_{2 j-1}\\
b_{ j \uparrow} &=&  \left(\prod_{i=1}^{j-1} \sigma^x_{2 i-1}\right)\sigma^y_{2 j-1}\\
a_{ j \downarrow} &=&  \left(\prod_{i=1}^{L} \sigma^x_{2 i -1}\right) \left(\prod_{i=j+1}^{L} \sigma^x_{2 i}\right) \sigma^y_{2 j}\\
b_{ j \downarrow} &=& -\left(\prod_{i=1}^{L} \sigma^x_{2 i-1}\right) \left(\prod_{i=j+1}^{L} \sigma^x_{2 i}\right) \sigma^z_{2 j}
\end{eqnarray}
\noindent followed by the Mattis-Nam transformation~\cite{Mattis72}
\begin{eqnarray}
(\sigma^x_{2j-1},\sigma^y_{2j-1},\sigma^z_{2j-1}) &=& (\tau^z_j,\tau^x_j\rho^x_j,\tau^y_j\rho^x_j) \\
(\sigma^x_{2j},\sigma^y_{2j},\sigma^z_{2j}) &=& (-\tau^z_j\rho^z_j,\tau^z_j\rho^y_j,\rho^x_j)
\end{eqnarray}
\noindent where $\tau^\alpha$ and $\rho^\alpha$ are Pauli matrices, the spin operators are mapped onto
\begin{eqnarray}
S^x_j &=& \frac{1}{4}\tau^x_j(1+\rho^z_j)\prod_{i=j+1}^L(-\rho^z_i)\\
S^y_j &=& \frac{1}{4}\tau^y_j(1+\rho^z_j)\prod_{i=j+1}^L(-\rho^z_i)\\
S^z_j &=& \frac{1}{4}\tau^z_j(1+\rho^z_j)
\end{eqnarray}
\noindent such that for a Hamiltonian built with spin operators the local parity operators $\rho^z_j=-(2 n_{j\uparrow}-1)(2 n_{j\downarrow}-1)$ are constants of motion. Thus, in general we will obtain a spin $1/2$ model coupled to a static $Z_2$ gauge field.
In our case, the Hamiltonian $H_0$ is mapped onto the spin model
\begin{eqnarray}
\nonumber H_0&=&-\frac{J}{16}\sum_{j=1}^{L-1}\left( (1+\gamma) \tau^x_j\tau^x_{j+1}+(1-\gamma)\tau^y_j\tau^y_{j+1} \right)\\
&& \times(1+\rho^z_j)(1+\rho^z_{j+1})
\end{eqnarray}
\noindent and the spin up and down parity symmetries $P_\uparrow=\prod_{i=1}^L \tau^z_i$ and $P_\downarrow=\prod_{i=1}^L (-\tau^z_i\rho^z_i)$. 
All the eigenstates of $H_0$ can be classified in terms of the eigenvalues $r_j$ of the operators  $\rho^z_j$, and the model can be exactly solvable in each of these subspaces. In details, for a configuration $\{r_j\}$ we can obtain a certain number of disjoint $XY$ chains, such that the ground state is achieved by the homogeneous configuration $\{r_j=1\}$. In the sector $\{r_j=1\}$ we get the $XY$ model~\cite{lieb61}
\begin{equation}
H_0=-\frac{J}{4}\sum_{j=1}^{L-1} (1+\gamma) \tau^x_j\tau^x_{j+1}+(1-\gamma)\tau^y_j\tau^y_{j+1}
\end{equation}
\noindent and by performing the Jordan-Wigner transformation
\begin{eqnarray}
a_j&=&\left(\prod_{i=1}^{j-1}\tau^z_i\right)\tau^x_j\\
b_j&=&\left(\prod_{i=1}^{j-1}\tau^z_i\right)\tau^y_j
\end{eqnarray}
\noindent where $a_j$ and $b_j$ are Majorana operators, we acquire the Kitaev chain at the half-filling
\begin{equation}
H_0=-i\frac{J}{4}\sum_{j=1}^{L-1} (1-\gamma)a_j b_{j+1} -(1+\gamma) b_j a_{j+1}
\end{equation}
\noindent For $\gamma \neq 0$, this model shows a gapped quantum phase and two Majorana modes which are decoupled at the thermodynamic limit and are localized at the edges of the chain~\cite{kitaev01}. The existence of the Majorana fermions is topologically protected by the symmetry $P_\uparrow = \prod_{i=1}^L(-i a_i b_i)$. In particular at the Ising point $\gamma=1$, these Majorana fermions are $a_1$ and $b_L$ which are expressed in terms of the fermions $a_{j \sigma}$ and $b_{j \sigma}$ as
\begin{eqnarray}
a_1 &=& -i a_{1\downarrow}b_{1 \uparrow}\left(\prod_{i=1,\sigma}^L i a_{i\sigma}b_{i\sigma}\right)\\
b_L &=& -a_{L \uparrow} b_{L \downarrow}\left(\prod_{i=1}^L i a_{i\uparrow}b_{i\uparrow}\right)
\end{eqnarray}
\noindent There are two degenerate ground-states which are the vacuum state $\ket{0}$, in this case defined by $i a_{j+1}b_j\ket{0}= \ket{0}$ for $j=1,\cdots,L-1$, and the first excited state $\ket{1}=\eta_0^\dagger\ket{0}$, where $\eta_0$ is the complex zero-mode which is a combination of the unpaired Majorana fermions, in this case $\eta_0=(a_1 + i b_L)/2$. Conversely, for $\gamma\neq 1$, the zero-mode has the one-particle energy $\Lambda_0\simeq 2/(1+\gamma)((1-\gamma)/(1+\gamma))^{L/2}$ as $L\to \infty$, such that in a finite system, the ground state $\ket{0}$ is
non-degenerate. In the thermodynamic limit $\Lambda_0\to 0$ exponentially and the ground-state remains two-fold degenerate until $\gamma\neq 0$, since the Majorana fermions are topologically protected. At the critical point $\gamma=0$, the quantum phase becomes gapless and without localized zero modes. We note that the zero-mode is bosonic because it involves a product of an even number of fermions $a_{j \sigma}$ and $b_{j \sigma}$, specifically by applying the parity transformation $P$ and $P_\uparrow$ we obtain $P\ket{1}=\ket{1}$ and $P_\uparrow \ket{1}=-\ket{1}$.
Due to the presence of the zero-mode, the quantum phase shows a spontaneous breaking of the $P_\uparrow$ symmetry with the presence of long-range order. The long-range order can be characterized by the end-to-end correlation function
\begin{equation}
\langle S^x_1 S^x_L \rangle = -\frac{1}{16}\left\langle \tau^x_1(1+\rho^z_1)\left(\prod_{i=2}^{L-1}(-\rho^z_i)\right) \tau^x_L (1+\rho^z_L)\right\rangle
\end{equation}
\noindent that tends to $\langle S^x_1 S^x_L \rangle = -(-1)^L\gamma/(1+\gamma)^2+\mathcal O(1/L)$ when the average is calculated with respect to the vacuum state. Furthermore, the entanglement spectrum in this ordered phase has at least twofold degeneracy. In this case, the density-matrix eigenstates transform in a nontrivial way under a projective representation of the symmetry.

On the other hand, the hopping term $H_1$ is mapped onto
\begin{eqnarray}
\nonumber H_1 &=& \frac{t}{2}\sum_{j=1}^{L-1}(\tau^x_j\tau^x_{j+1}+\tau^y_j\tau^y_{j+1}+1)\rho^x_j\rho^x_{j+1}\\
 && + \tau^z_j\tau^z_{j+1}\rho^y_j\rho^y_{j+1}
\end{eqnarray}
\noindent which breaks the local parity symmetry $\rho^z_j$ and shows a trivial gapless quantum phase.
For small $t$, we can describe the influence of $H_1$ in perturbation theory through an effective Hamiltonian~\cite{sachdev}. By considering the Ising point $\gamma=1$ for simplicity, in the limit $t\to 0$ there are two degenerate ground-states $\ket{\Rightarrow}\ket{\Uparrow}=\ket{\rightarrow \rightarrow \cdots \rightarrow}\ket{\uparrow \uparrow \cdots \uparrow}$ and $\ket{\Leftarrow} \ket{\Uparrow}=\ket{\leftarrow \leftarrow \cdots \leftarrow}\ket{\uparrow \uparrow \cdots \uparrow}$. For small $t$ the ground-state energy is $E_0=-(J/2+4t^2/(3J))(L-1)+\mathcal O(t^3)$ which remains two-fold degenerate. 
Conversely, for $t\to 0$ the low-lying excited states are $\ket{\phi_i}\ket{\Uparrow}=\ket{\rightarrow \cdots \rightarrow \leftarrow \cdots\leftarrow}\ket{\Uparrow}$ and $\ket{\psi_i}\ket{\Uparrow}=\ket{\leftarrow  \cdots \leftarrow \rightarrow \cdots\rightarrow}\ket{\Uparrow}$, where the kink is located between sites $i$ and $i + 1$. 
By performing a straightforward calculation, we achieve the effective Hamiltonian in this basis, $H_{eff} = const + 4 t^2/J \sum(\ket{\phi_{i}}\bra{\phi_{i+2}}+\ket{\psi_{i}}\bra{\psi_{i+2}}+h.c.)\ket{\Uparrow}\bra{\Uparrow} $, such that at the second order the term $H_1$ moves a kink to its next-nearest-neighbor sites. The Hamiltonian $H_{eff}$ is therefore diagonalized by going to the momentum space basis and the one-particle eigenstates have energies $\epsilon_k = const + 8 t^2/J \cos(2k)$.

\section{\label{sec.qpt} Phase diagram}
\begin{figure}
[h!]
\includegraphics[width=0.76\columnwidth]{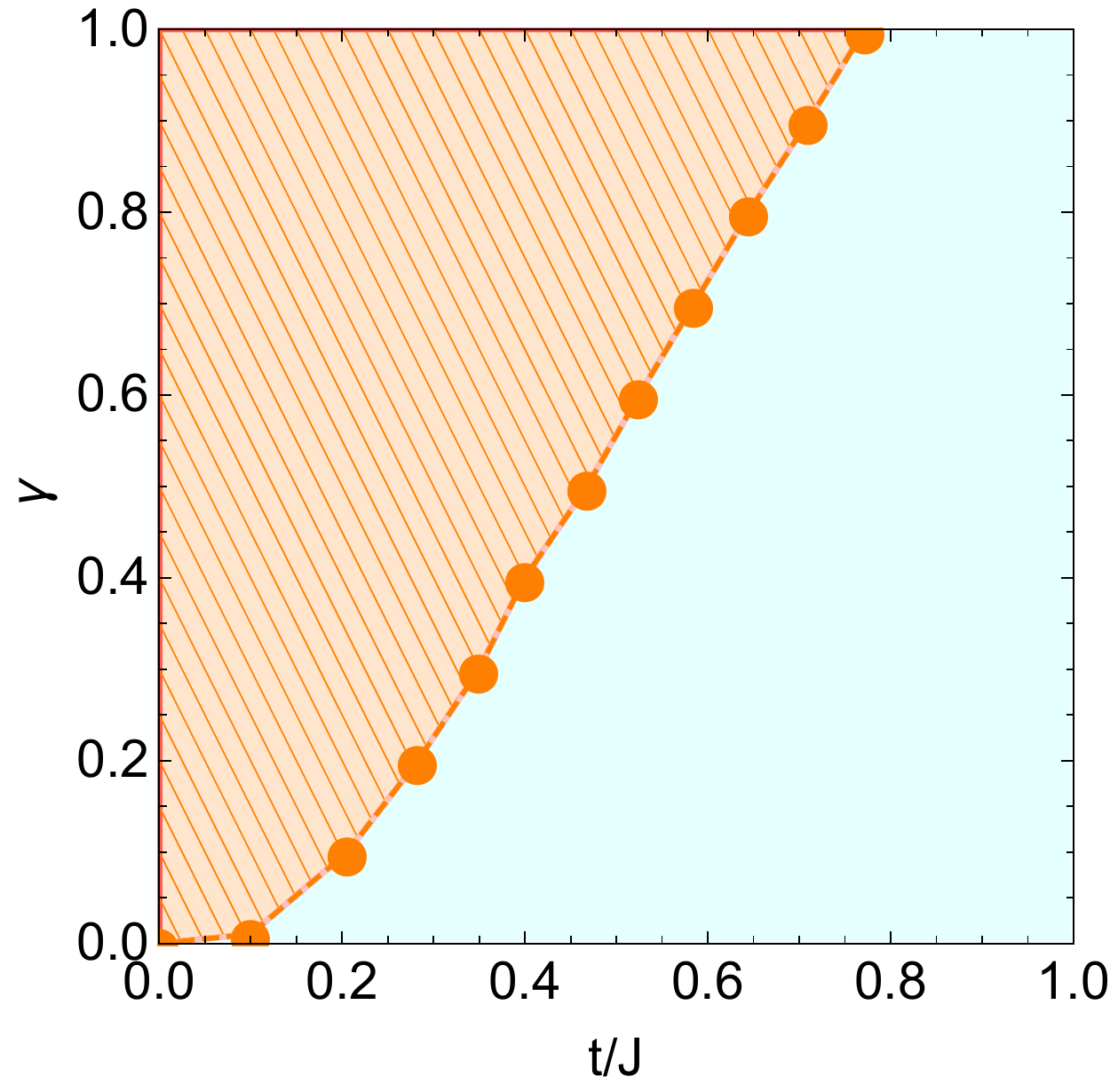}
\caption{ The phase diagram of the model in function of the hopping amplitude $t$ and the anisotropy $\gamma$. The orange marked with lines and the cyan regions correspond to the topological non-trivial and trivial phases, respectively. The dots represent the critical points $t=t_c(\gamma)$ which are calculated by doing a finite-size scaling analysis as illustrated in Fig.~\ref{fig:finitesize}.
}
\label{fig:dia}
\end{figure}

The Hamiltonian $H_1$ drives the system toward a trivial phase without zero modes. Specifically, the two phases are separated by a second order quantum phase transition point~\cite{sachdev}. We estimate the phase diagram in Fig.~\ref{fig:dia} by doing a finite-size scaling study of the end-to-end correlation $\langle S^x_1 S^x_L \rangle$, which in the limit $L\to \infty$ is non-zero only in the topological phase.
\begin{figure}
[h!]
\includegraphics[width=0.76\columnwidth]{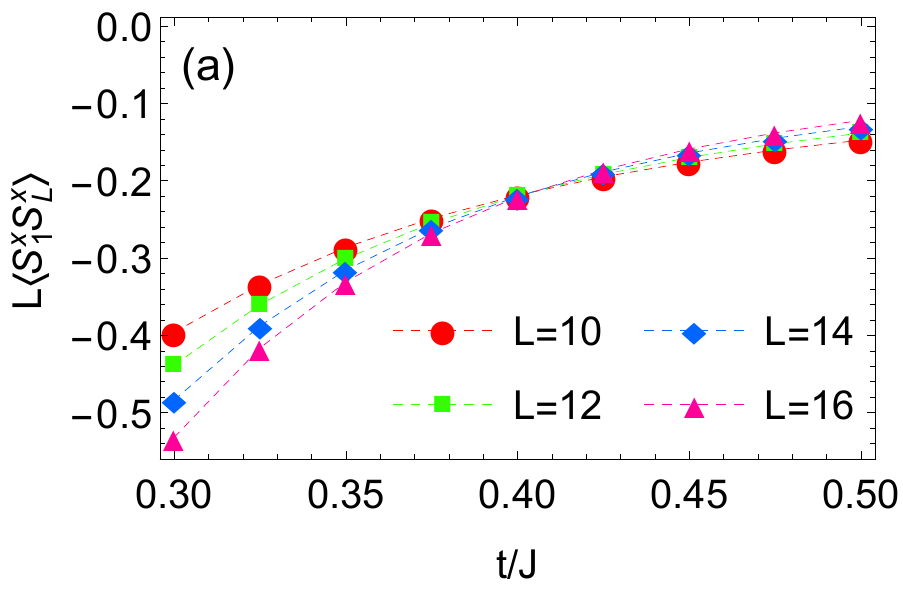}
\includegraphics[width=0.76\columnwidth]{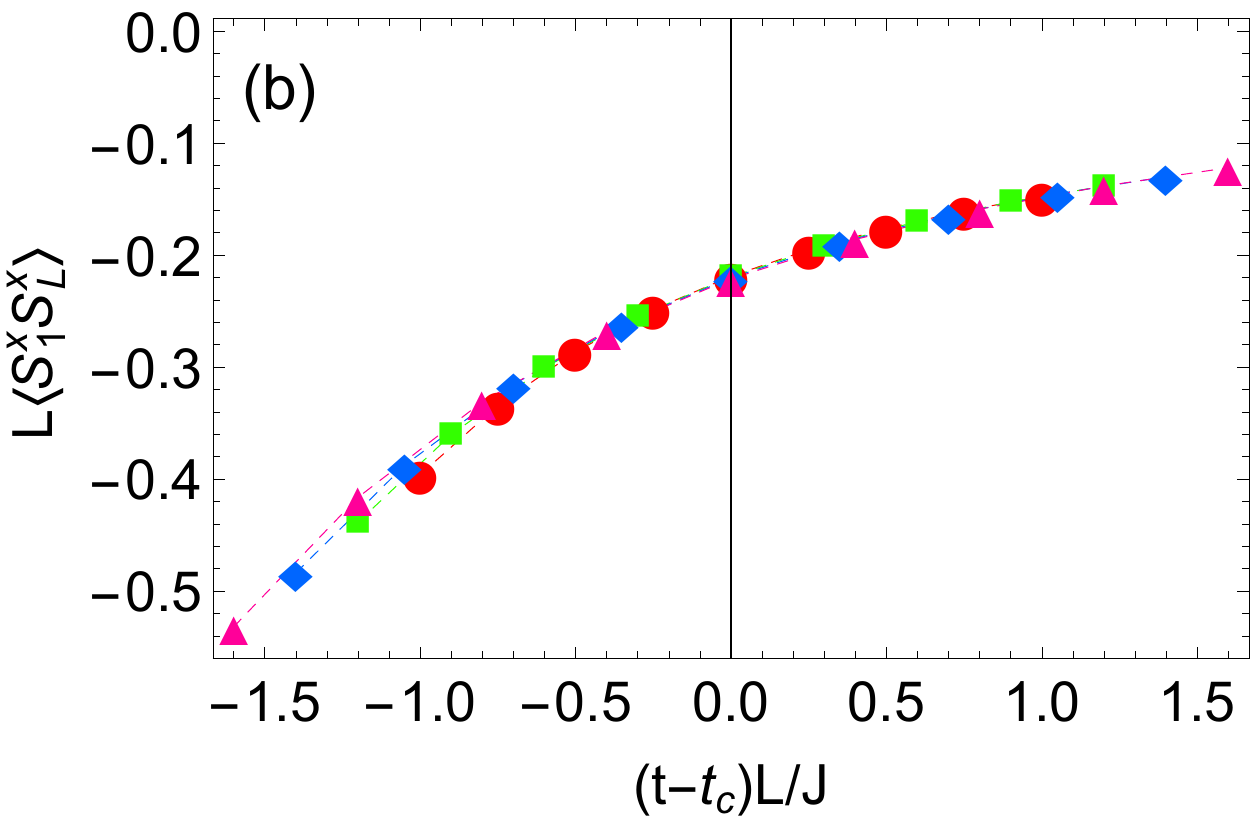}
\caption{ The scaling plots of the end-to-end correlation function at $\gamma=0.4$. The crossing at $t = t_c \approx 0.4 J$ in (a) and the scaling collapse in (b)
yield the critical exponents $\beta=1/2$ and $\nu=1$.
}
\label{fig:finitesize}
\end{figure}
In details, we change the hopping amplitude $t$ at fixed anisotropy $\gamma$ for finding the phase transition point $t=t_c(\gamma)$. In order to do the analysis, we find the optimal approximation to the ground state by performing a variational search in the matrix product state space~\cite{schollwock11}. The finite-size scaling ansatz of the correlation function reads $\langle S^x_1 S^x_L \rangle=L^{-2\beta/\nu} S((t-t_c)L^{1/\nu})$ where $\beta$ and $\nu$ are the critical exponents describing the singular behaviors of the end-to-end correlation and the correlation length as $ \langle S^x_1 S^x_L \rangle \sim (t-t_c)^{2\beta}$ and $\xi \sim \abs{t-t_c}^\nu$, respectively. As expected from the finite-size scaling ansatz, $ L^{2\beta/\nu}\langle S^x_1 S^x_L \rangle$ for different sizes cross at $t_c$ with $\beta/\nu=1/2$ (see Fig.~\ref{fig:finitesize}(a)).
Furthermore, when we plot $ L^{2\beta/\nu} \langle S^x_1 S^x_L \rangle$ versus $(t-t_c)L^{1/\nu}$, data points collapse into a single curve with $\nu=1$, as shown in Fig.~\ref{fig:finitesize}(b), such that we estimate the exponents $\beta=1/2$ and $\nu=1$, in agreement with the 2D classical Ising universality class.


\section{\label{sec.conclusion} Conclusion}
In summary, we have investigated the quantum phases of a chain of spinful fermions in the presence of a $XY$ antiferromagnetic interaction. We have shown how the $XY$ term can be exactly mapped onto a Kitaev chain at half-filling. Thus, the model emulates a  topological quantum phase with Majorana fermions, which can be employed for quantum computing purposes. The effects of the nearest hopping are analysed with the help of the perturbation theory and numerically by calculating the ground state via a matrix product state variational method. In particular the standard finite-size scaling method has been applied to determinate the phase diagram of the model. 

\end{document}